\documentclass{article}
\usepackage[utf8]{inputenc}
\usepackage{datetime}
\usepackage{authblk}
\usepackage{setspace}
\usepackage[margin=1.25in]{geometry}
\usepackage{graphicx, animate}
\graphicspath{ {./figures/} }
\usepackage{booktabs}
\usepackage{subcaption}
\usepackage[flushleft]{threeparttable}
\usepackage{amsmath}
\usepackage[  
    colorlinks=true,
    linkcolor=red
]{hyperref}

\usepackage[style=ieee, 
citestyle=numeric-comp,
sorting=none]{biblatex}
\title{Friday the 13th Hailstorm in the province of Bulacan, Philippines (13 August 2021): A Case Study}

\author[1,2*]{Generich H. Capuli}

\affil[1]{Project Severe Weather Archive of the Philippines}
\affil[2]{Department of Earth and Space Sciences, Rizal Technological University, Brgy. Malamig, Mandaluyong City, Metro Manila 1550, Philippines}
\affil[*]{Corresponding author. Email: genhcapuli@rtu.edu.ph}

\date{Friday 13$^{\text{th}}$ December, 2024}

\begin{document}

\maketitle

%%%%%% Abstract %%%%%%
\begin{abstract}
This case study presents a thorough investigation of the environmental setup that led to the hail-producing severe storm that impacted the municipality of Norzagaray and City of San Jose Del Monte, including other nearby areas, in the province of Bulacan on the afternoon of August 13, 2021. During this period, 2-5 cm and potentially as large as $\sim$8 cm diameter hail was reported over these locations of Bulacan. For this purpose, the combination of HIMAWARI-8 AHI, PLDN and its flash counts, and meteorological indices; synoptic, thermodynamic, and kinematic indices, calculated from the ERA5 reanalysis are utilized to understand the nature of the hail event. In the morning, the pre-convective environment was comprised by a warm inversion layer that inhibited storm initiation, until the arrival of ample moisture and convective heating in the afternoon. By the afternoon, model sounding analysis revealed that the environment transitioned into uncapped profile with steep low-level lapse rate owing to warm, moist south-westerly wind flow from the Manila Bay in the lower troposphere and north-easterlies aloft crossing the SMMR induced by a weak low-pressure system located in the eastern Philippine Sea, with minimal turning on the wind profile. This promoted low-level convergence within the area of interest and build up of instability. The updraft associated with convectively unstable atmosphere, sufficient cloud-layer bulk shear, and storm nudging at its maturing phase countered entrainment-driven dilution and aided the growth of ice crystals by rapid collection of supercooled cloud liquid particles, which ultimately led to formation of hailstones.
\end{abstract}

%%%%%% Main Text %%%%%%

\section{Introduction}

Severe thunderstorms are one of the most costly natural hazards \cite{1,2,3}. They can lead to straight-line winds, intense rainfall, tornadoes and in particular, large hail. These hydrometeors are ice pellets that forms inside these systems and grows at an altitude of 4800-6400 m above the ground, where temperatures are below -0 to -20 $^{\circ}$C. Water vapor in the air moves up with the air parcel inside the storm and after condensing, turns into ice crystals \cite{4}. Most hailstorms are caused by thunderstorms with lightning; however, only 60\% of thunderstorms are accompanied by hail \cite{5}. An intricate balance of instability and vertical wind shear, even external features, should exist for hail embryos to form and maintain along the storm's updraft \cite{6,7}. In some thunderstorms, the hail hits the hot air as it descends to the ground and turns into water droplets; that is why in areas with hot climates, hail does not reach the ground \cite{8}. Regardless, they still occur especially for tropical climates such as in the Philippines.  

Hail has been studied for more than half a century globally, particularly started through observations of Browning and Ludlam \cite{8a} in a severe hailstorm in Wokingham, England, by employing five radars, and found some distinct radar characteristics. Yet, relatively little is known about the storms that produce hail or the environments that support hail-producing storms in the Philippines. Hail research in the archipelago goes back near the end of 1920s by Selga \cite{9}. Hail observations were extremely limited at this time, but some frequency of hail was noted near the Northern Luzon and Greater Metro Manila were subject to such hailstorm activity. In fact, hail reports within these areas are some of the highest across the archipelago and is the hotspot for such activity according to recent severe weather climatology constructed by Severe Weather Archive of the Philippines \cite[SWAP;][]{10}.

Furthermore, it was recently found out that severe weather activity, including Hail events, in the Philippines are found to have an annual cycle similar to the United States, with the bulk of hailstorm activity occurring during the spring and summer; hot, dry season from March to May extending to the Southwest Monsoon period \cite{11,10}. These tend to occur in the afternoon and early evening - a typical diurnal pattern across the tropics and in the archipelago \cite{12}, which are aligned with the observations of Selga \cite{9}, who suggested that most hail events in the Philippines occur during the dry season. It was also seen that orographic effects can compensate for weakly-sheared environments that generate hailstorms in the European continent \cite{13}. And it can be same given that the Equator's horizontal temperature gradient is very small, and upper level air circulation is comparatively weak. Although, current research suggest that the fraction of severe weather environments have relatively increased within the tropics, particularly in ITCZ, modulated by the increase of deep-layer shear over the past 4 decades \cite{14}.   

While we understand some parts of thunderstorm and hail formation \cite{15,16}, research gaps persist in our understanding of the dynamics and microphysical processes in local thunderstorm evolution, intensification, and large hail formation, and these gaps limit our forecasting capabilities. Much of our process understanding is based on studies of thunderstorms in the United States, most notably the Great Plains \cite[e.g.][]{17,18,19,20,21}, where atmospheric characteristics differ significantly from Southeast Asian countries such as Philippines, as convective available potential energy (CAPE) are typically lower, vertical wind shear are much higher, and local orographic influence is noticeable in form of precipitation events\footnote[1]{Also known as Orographic Precipitation. Terrain effects were also seen to affect Tropical Cyclone rainfall and prevailing wind patterns \cite[][, and references therein]{ft1,ft2,ft3,ft4}.} \cite{22,23}.

In addition, anthropogenic climate change (ACC) is expected to have significant impacts on hailstorms, which can result in damaging consequences \cite{24,25}. The quality of low-level moisture that results to convective instability \cite{26,27,28}, melting level height \cite{29,30}, and vertical wind shear, particularly deep-layer shear \cite[DLS,][]{31,32} are all expected to change with geographical variability, leading to changes in hailstorm frequency and severity \cite{33,34,35}. However, due to limited direct observations of hail, incomplete understanding of the microphysical and dynamical processes, the response of hailstorms to warming remains highly uncertain \cite{15,24}.

Hailstorm case studies highlight the influence of mesoscale environments across various regions. Some of the case studies by Huang et al. \cite[][18 June 2013 - China]{36}, Bechis et al. \cite[][26 November 2018 - Argentina]{37} during the RELAMPAGO\footnote[2]{Remote sensing of Electrification, Lightning, And Mesoscale/microscale Processes with Adaptive Ground Observations} and CACTI\footnote[3]{Clouds, Aerosols, and Complex Terrain Interactions} field campaigns, and Tai et al. \cite[][2 July 1998 - Taiwan]{38} all asserted that their respective and selected hailstorm events are by-product of favorable synoptic forcing (i.e. frontal system, upper-level vortex, jet streak orientation), conditional instability, and ambient vertical shear, including the unique topography such as mountain regions that produces orographically-induced vertical lifting, for convective initiation that led to sustained supercells and/or other mesoscale systems causing significant severe weather events (SWEs). Throughout the years, such environmental setups are pre-dominantly found in U.S. Great Plains \cite{39} and in different European regions \cite[][, and references therein]{40}.

\begin{figure}[!ht]
    \centering
    \includegraphics[width=\textwidth]{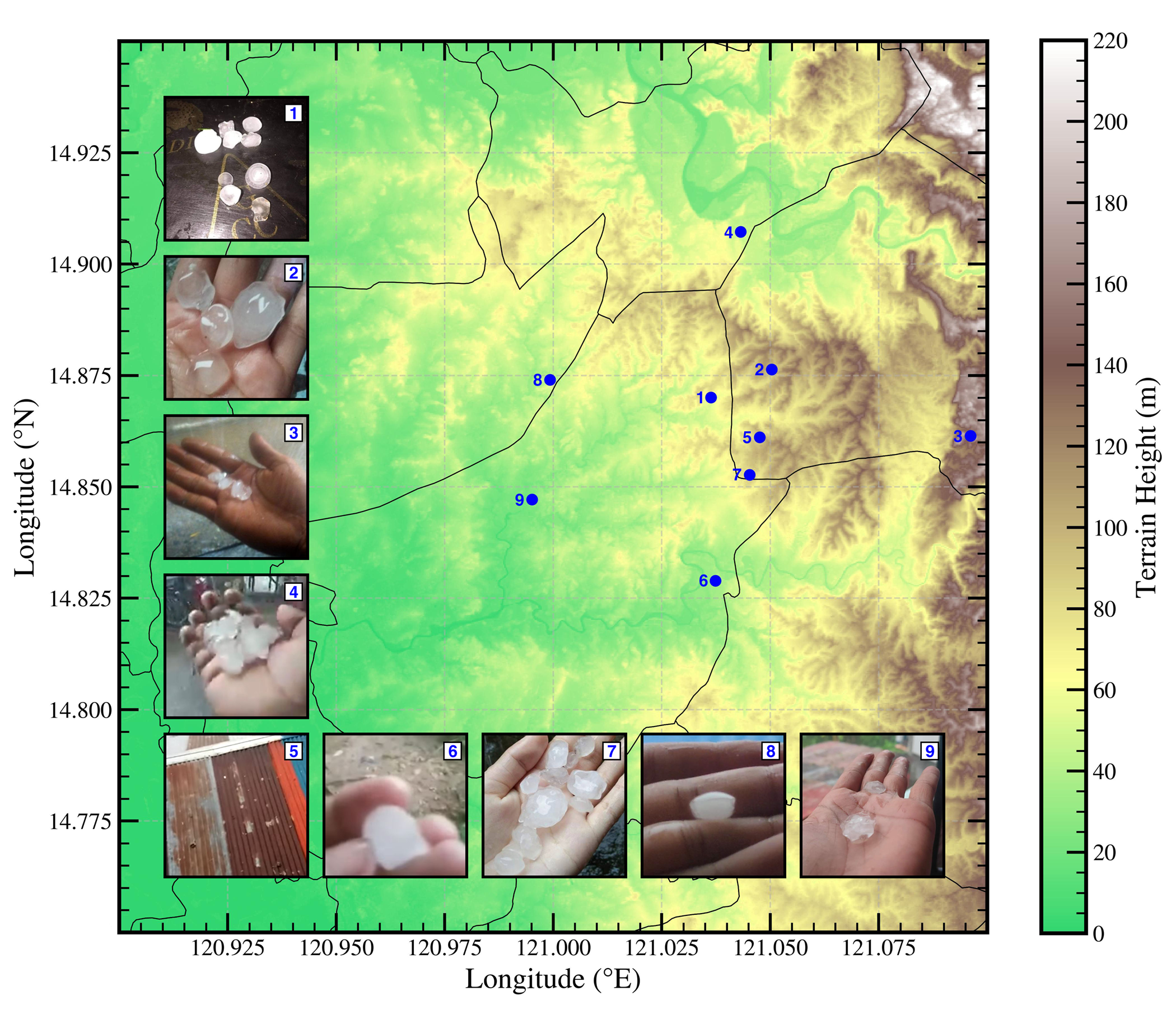}
    \caption{Various citizen hail reports in the municipalities of Norzagaray, San Jose Del Monte, and Sta. Maria, Bulacan. Citizen hail samples are included.}
    \label{fig:1}
\end{figure}

On the afternoon of August 13, 2021, a significant hail was observed, motivating its selection as a case study. Thunderstorms erupted in the eastern section of Bulacan province at around 07 UTC bringing heavy rains and swath of hail in parts of City of San Jose Del Monte, and both municipalities of Norzagaray and Sta. Maria, Bulacan. Hail sample estimated size of $>$ 2-5 cm in diameter, with a potential hail sample having an estimated size of 8 cm; by far the largest hail recorded in the Philippines (particularly in Barangay Tigbe, Norzagaray), was reported from different sources: social networks and citizen reports. We dubbed this event as the 'Friday the 13th Hailstorm' in Bulacan given that it occurred on 'Friday the 13th' synonymous to the superstitious belief of having a bad luck or spooky day. As to why we selected this case to uncover why it was 'spooky' i.e. the environment capable of producing such severe storm and hazards. In a serious note, these reports are scattered across the social media networks and has been archived/compiled well in Project SWAP. The distribution of the hail reports is depicted in Figure \ref{fig:1}.  

The aim of this paper is to characterize the modifications in the pre-convective environment that lead to the initiation of this severe hailstorm utilizing the available reanalysis dataset. We also seek to characterize the different stages of the storm life-cycle and their association with the severity of hail, by combining remote sensing tools and the surface hail reports. Despite the absence of radar data, lightning data is available from PAGASA Lightning Data Networks during that time. 

The present paper is segmented as follows; Section 2 is the overview of the materials and methods, particularly pertaining to the reanalysis dataset, profile parameters, satellite and lightning data, utilized in this study. The environmental analysis of synoptic scale, mesoscale, and storm dynamics of the selected hailstorm event is discussed in Section 3. Finally, a summary and further discussion in Section 4.

\section{Materials and Methods}

\subsection{Environmental Reanalysis Dataset}

This research was complemented by analyzing potentially severe convective environment associated with the 13th of August 2021 Bulacan Hailstorm event. Reanalysis datasets allow identifying the potential development of severe convective storms. In addition, reanalysis data have also been used in the computation of instability indices for convective forecasting and hazardous convective weather such as tornadoes, wind-driven events, and large hail \cite{23,41}.

\begin{table}[h!t!]
\small
\caption{Environmental Sounding Parameters}
    \centering
    \begin{tabular}{llll}
    \hline\hline
    Parameter & Definition & Notes & Units \\
    \hline
    CAPE & Convective Available Potential Energy & MU parcel & J kg$^{-1}$ \\
    CAPE$_{\text{HGZ}}$ & CAPE between 0 $^{\circ}$C and -20 $^{\circ}$C & MU parcel & J kg$^{-1}$ \\
    CAPE$_{\text{$<$HGZ}}$ & CAPE below 0 $^{\circ}$C & Proxy: CAPE$_{06}$; MU parcel & J kg$^{-1}$ \\
    ECAPE & Entraining CAPE & MU parcel & J kg$^{-1}$ \\
    DCAPE & Downdraft CAPE & & J kg$^{-1}$ \\
    CIN & Convective Inhibition & SB parcel & J kg$^{-1}$ \\
    LCL & Lifted Condensation Level & SB parcel & m \\
    LFC & Level of Free Convection & MU parcel & m \\
    LR$_{03}$ & 0-3 km Lapse Rate & Low-level Lapse Rate & $^{\circ}$C km$^{-1}$ \\
    LR$_{36}$ & 3-6 km Lapse Rate & Mid-level Lapse Rate & $^{\circ}$C km$^{-1}$ \\
    RH$_{13}$ & Mean 1-3 km Relative Humidity & & frac / $\%$\\
    RH$_{16}$ & Mean 1-6 km Relative Humidity & & frac / $\%$\\
    SHR$_{01}$ & 0-1 km Bulk Wind Difference & Low-level Shear (LLS) & kt / m s$^{-1}$ \\
    SHR$_{06}$ & 0-6 km Bulk Wind Difference & Deep-layer Shear (DLS) & kt / m s$^{-1}$ \\
    SHR$_{\text{LCL-EL}}$ & Cloud-layer Shear & Proxy: 200-900 hPa BWD & kt / m s$^{-1}$ \\
    SRW$_{01}$ & Mean 0-1 km Storm-relative Winds & Using Bunker's RM & kt / m s$^{-1}$ \\
    WMAXSHEAR & Undiluted Updraft Velocity $\times$ SHR$_{06}$ & MU parcel & m$^{2}$ s$^{-2}$ \\
    WMAXSHEAR$_{06}$ & 0-6 km Updraft Velocity $\times$ SHR$_{06}$ & MU parcel & m$^{2}$ s$^{-2}$ \\
    FZL & Freezing level (0 $^{\circ}$C) & MU parcel & m \\
    HGZ & Hail Growth Zone (0 $^{\circ}$C and -20 $^{\circ}$C) & MU parcel & m \\
    \hline
    \end{tabular}
    \label{table:1}
\end{table}

Environmental reanalysis data for this case were obtained from the fifth-generation ECMWF reanalysis or ERA5 reanalysis through Climate Data Store \cite[CDS,][]{42}. This reanalysis has been shown to faithfully represent the vertical profiles of severe convective events, particularly in the United States and Europe \cite{43,44,45}, but still, some known biases exist. In particular, some of the largest differences between ERA5 and observed soundings have been found in the boundary layer, low-level parcel and vertical shear parameters, especially in proximity to surface boundaries \cite{44,46}. Furthermore, these biases differ per geographic location and surface elevation. Fortunately, however, ERA5 remains as one of the most reliable and affordable reanalyses for understanding potentially severe convective environments, even in a global context \cite{14,43,44}. 

For large synoptic scale, the selected area extends 4$^{\circ}$ to 22$^{\circ}$ N and 110$^{\circ}$ to 135$^{\circ}$ E, while the mesoscale sector (including for the satellite visualization) spans from 13$^{\circ}$ to 19$^{\circ}$ N and 118$^{\circ}$ to 125$^{\circ}$ E both with a spatial resolution of 0.25$^{\circ}$ x 0.25$^{\circ}$. Hourly temporal allowed for a more detailed analysis of the hail event that occurred at around 07 UTC. This granularity enabled the examination of weather patterns leading up to, during, and after the significant hail event. For this case, the time steps of 06, 07, and 08 UTC where selected.

\subsection{Vertical Profile}

Observed proximity sounding (within 200 km of the event) from University of Wyoming (UWyo) database was also queried; specifically, the PAGASA Synoptic Station located 614 mASL at Tanay, Rizal station (WMO ID: 98433 - 14.57$^{\circ}$ N, 121.37$^{\circ}$ E). This rawinsonde observation (RAOB) was conducted routinely in a sub-daily basis: 00 UTC and 12 UTC. In addition, a sounding profile was also extracted from the combination of single-level and 137 hybrid sigma-pressure level ERA5 data. These soundings are implemented in the study to delineate the convective and kinematic environment of the significant event. 

The parcel profile associated in these soundings assumes a non-entraining, irreversible adiabatic ascent as recently described and developed by Peters et al. \cite{47} in Equation 25. Compared to pseudoadiabatic ascent wherein all condensate is assumed to fall out of an air parcel immediately as it forms \cite{48}, the adiabatic parcel profile now accounts the layer of mixed-phase condensate in which liquid and ice are present just below the triple point temperature based from conservation of energy as its initial condition, rather than conservation of moist entropy, which in turn reduces the errors that arises from using other parcel techniques. In fact, it was hinted by Xu and Emanuel \cite{49} and their analyses of RAOBs in the tropics that environmental temperature more closely resembles that of an adiabatic parcel method rather than pseudoadiabatic and can be more relevant to deep convection. 

Additionally, an entraining, irreversible adiabatic ascent was also added to the mix that allows for computation of the Entraining CAPE \cite[ECAPE;][]{50}. ECAPE accounts for the entrainment-dilution of the thunderstorm's updraft plumes due to the lack of storm-relative inflow and/or tropospheric dryness. Thus, providing a better picture of updraft intensity than the conventional, undiluted CAPE when utilized on this hazardous convective weather that also depends on middle- to upper-level vertical velocities. The undiluted buoyancy profile is calculated using the virtual temperature correction. Either most-unstable (MU) and surface-based (SB) parcel profiles depending on their meteorological significance were utilized across the thermodynamic parameters. CAPE and LFC were all assume the MU parcel, while convective inhibition (CIN) and lifted condensation level (LCL) both assume the SB parcel, since this parcel best detects the degree of low-level stability, and the lowest potential cloud base, respectively, in both surface-based and elevated storm scenario. 

Kinematically, storm-relative winds was subject to assumptions on storm motion since neither observed storm motions, supercell type (right-moving or left-moving) nor storm mode (supercell, multicell cluster, squall line, etc.) were attained. Thus, estimated storm motion were calculated using the Bunkers ID Method \cite[B2K; ][]{51}, also an important component of ECAPE. For northern hemisphere countries like the Philippines, the right-moving vector was assumed (storm moving towards the right of non-pressure weighted mean wind). Other kinematics such as vertical wind shear parameters (i.e. 0-1, 0-6 km AGL, LCL-EL shear) were computed as is throughout the hodograph. All soundings associated in this event were rigorously analyzed using SounderPy by Gillett \cite{52}. A list of meteorological parameters and their meanings were presented in Table \ref{table:1}.

\subsection{HIMAWARI-8 Satellite Data}

Data with a high spatial resolution and high temporal resolution from the Advance Himawari Imager (AHI) instrument aboard the HIMAWARI-8 \cite{53} were also used to distinguish the convective system associated with the significant hail event. HIMAWARI-8 comprises of 3 VIS bands (central $\lambda$ ranging from 0.47 {\textmu}m to 0.64 {\textmu}m), 3 NIR bands (central $\lambda$ ranging from 0.86 {\textmu}m to 2.3 {\textmu}m), and 10 IR bands (central $\lambda$ ranging from 3.9 {\textmu}m to 13.3 {\textmu}m). Particularly, this study selected the visible bands and infrared (IR; 10.4 {\textmu}m) wavelength derived from the atmospheric window band (Band 13; B13) to monitor infrared radiation from cold cloud-top [Brightness Temperature (BT) $<$ 235 K ($\sim$ -40 $^{\circ}$C)], free from interference caused by water vapor \cite{54}. Band 13 is also widely used on cloud identification and predicting extreme rain events with small lag time \cite{55}. The visible bands has a spatial resolution of 500 m, while the IR band has a spatial resolution of 2 km. 

In addition, we also used and constructed the Japan Meteorological Agency's (JMA) Daytime Convective RGB product for visualizing convective cloud characteristics that are key to understanding severe weather phenomena, particularly hail. By focusing on specific color channels sensitive to cloud phase and temperature, we can identify areas where ice particles, whether large and small, are likely present within the convective storm. Particularly, it takes the difference between the temperature and radiance of selected bands; B10-B08 for the Red profile, B13-B07 for the Green profile, and B03-B05 for the Blue profile. All satellite channels used in the study was assimilated of every-10-minute full disk scan observations.

Penetrative overshooting top (OT) signatures and their relationship with hail severity is evaluated. OT’s are local minima in the observed IR cloud top temperature, and their presence is indicative of a deep convective updraft strong enough to penetrate through the equilibrium level (EL) and into the lower stratosphere, and tend to be associated with severe weather \cite{56}.

\subsection{PAGASA Lightning Network}

Lightning data was also requested through PAGASA and its Lightning Detection Network (hereafter, PLDN). PLDN encompasses 28 strategically positioned lightning sensors across extensive geographical areas in the archipelago, primarily stationed within PAGASA observing offices in efforts to strengthen early detection of thunderstorms, including its severe counterparts. Commenced in 2018, it has the capabilities to detect both Intra-cloud (IC) and Cloud-to-Ground (CG) lightning strikes through its real-time lightning tracking and warning functionalities providing the opportunity to identify rapid increases in lightning activity, known as lightning jumps, that often precede the development of severe weather \cite{57}. Thankfully, the lightning data for this case, starting from 06 to 08 UTC, are within the scope of the temporal availability (between 2020 to 2023). 

\section{Results}

\subsection{Overview of the Synoptic Setup}

\begin{figure}[!ht]
    \centering
    \includegraphics[width=\textwidth]{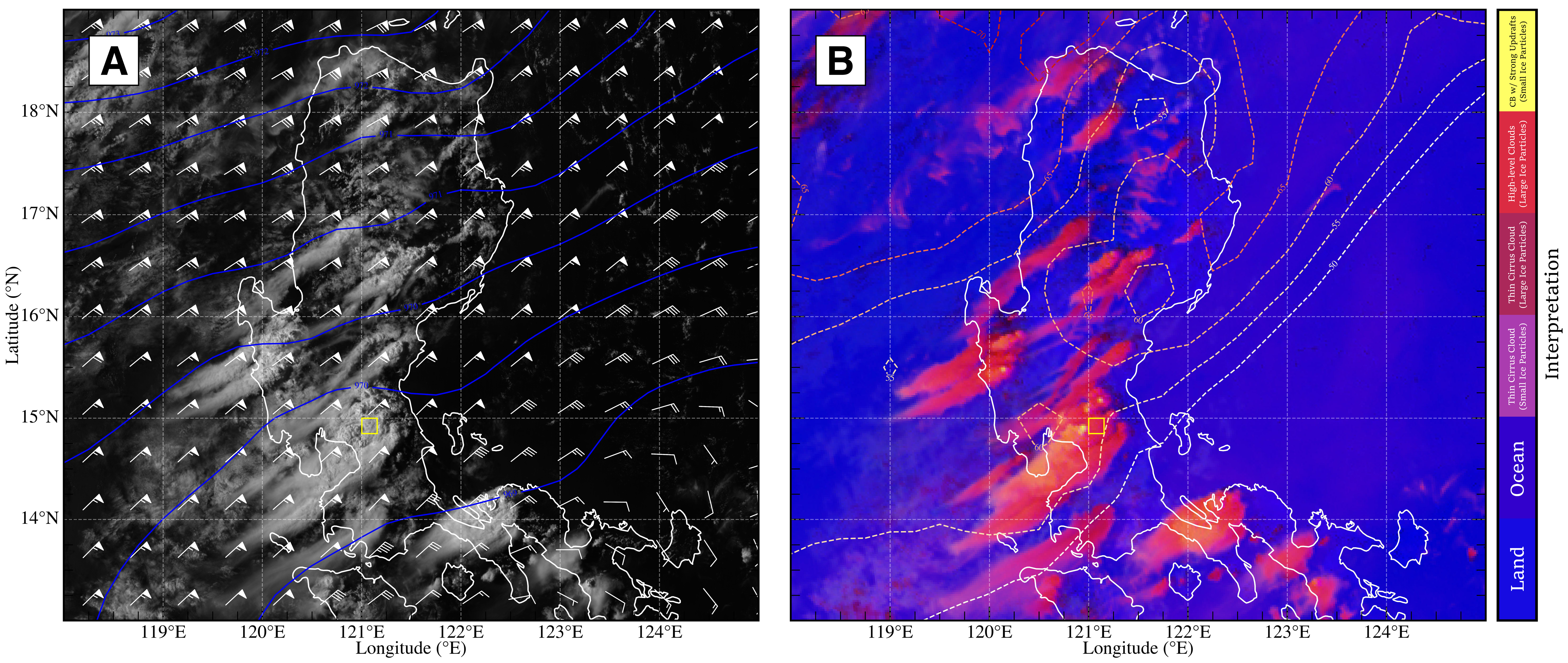}
    \caption{HIMAWARI-8 AHI scans at 07 UTC; (a) Band 03 0.64 {\textmu}m with Cloud-layer Shear Vectors (kt) and 300-hPa Geopotential heights (blue; dam), and (b) Daytime Convective RGB with Cloud-layer Shear Magnitudes $>$ 50 kts. The area of interest is demarcated in yellow box.}
    \label{fig:2}
\end{figure}

At 07 UTC of August 13, 2021, a semi-discrete thunderstorm was evident in the HIMAWARI-8 AHI scan along the Bulacan province with pronounced OT near the vicinity where the hail reports were identified (Fig. \ref{fig:2}a). Close inspection of this convective system using Daytime Convection RGB imagery suggest vigorous updrafts with small ice particles, while larger hydrometeors can be denoted along the thick high-level clouds close to the area of interest where the various hail reports are seen (Fig. \ref{fig:2}b). Within the updraft of this storm, supercooled water droplets and ice both coexisted in the upper levels aided by the strong buoyancy and 30 m s$^{-1}$ ($\sim$60 kt) of cloud-layer shear magnitude. 

\begin{figure}[!ht]
    \centering
    \includegraphics[width=\textwidth]{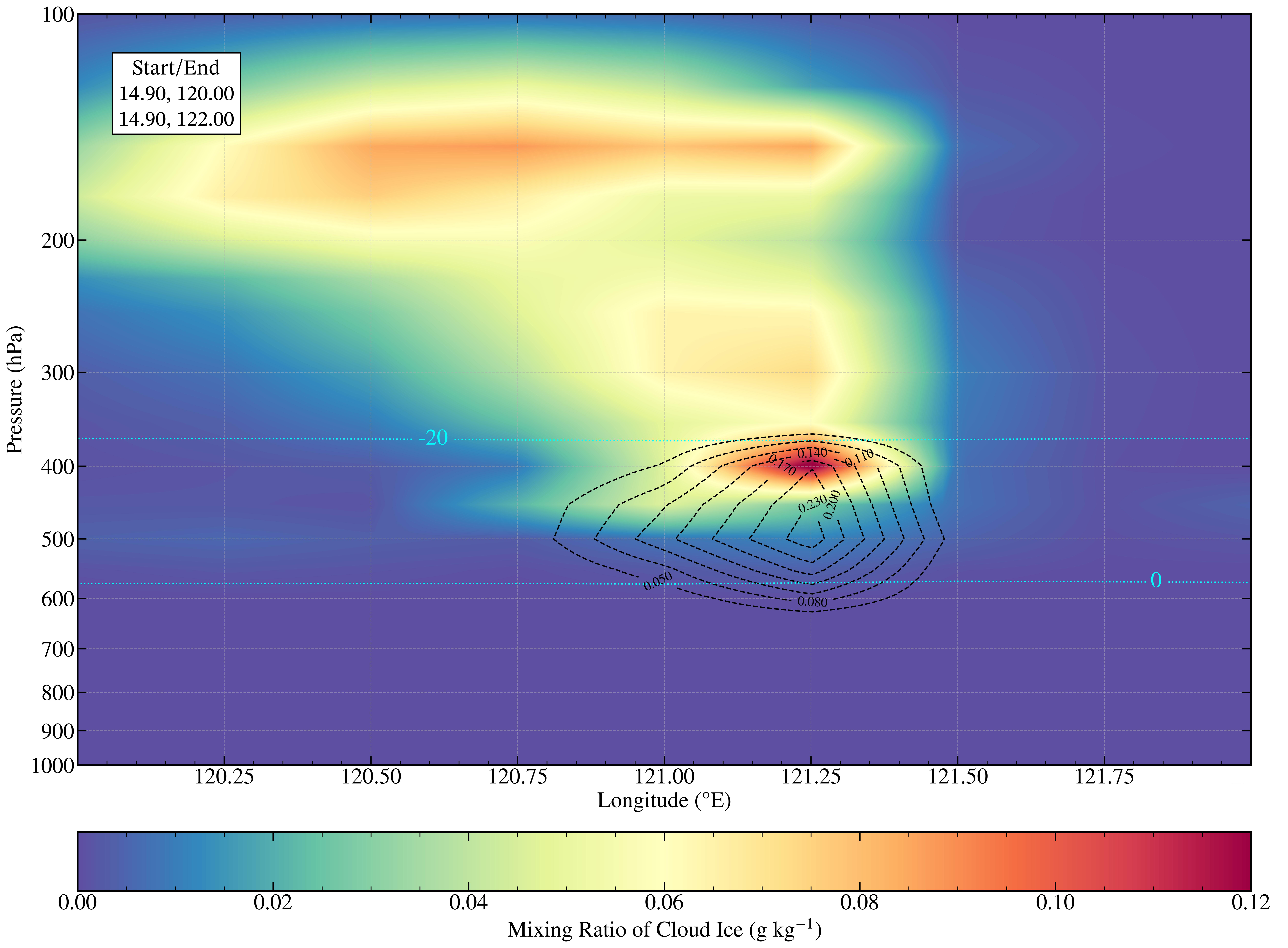}
    \caption{07 UTC ERA5 Vertical Cross Section of Cloud Ice (shaded) and Liquid Mixing Ratios (dashed; g kg$^{-1}$). The FZL; denoted as the 0 $^{\circ}$C isotherm, and HGZ is also identified as blue dotted lines.}
    \label{fig:3}
\end{figure}

\begin{figure}[!ht]
    \centering
    \includegraphics[width=\textwidth]{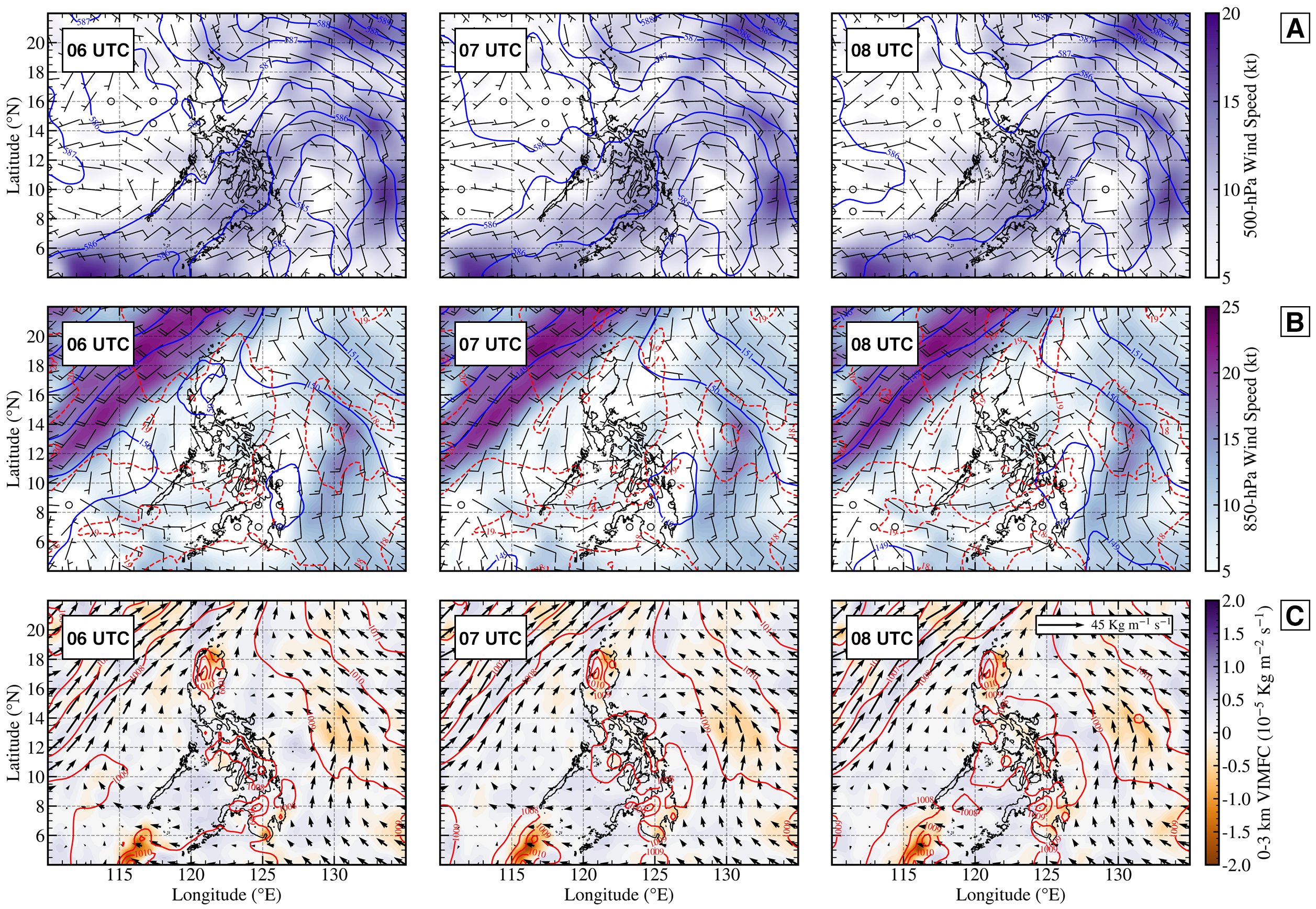}
    \caption{Synoptic environment of the Philippine archipelago before (06 UTC), during (07 UTC), and after (08 UTC) the event. (a) 500-hPa Geopotential heights (blue; dam) and winds (kt). Contours are wind speeds $>$ 5 kt. (b) 850-hPa Geopotential heights (blue; dam), temperature (red; $^{\circ}$C), and winds (kt). Contours are wind speeds $>$ 5 kt. (c) Mean Sea Level Pressure (MSLP; red; hPa), Vertical Integrated Moisture Flux Convergence within the first 3 km layer (VIMFC; Kg m$^{-2}$ s$^{-1}$), and its wind components (quivers).}
    \label{fig:4}
\end{figure}

To support the HIMAWARI-8 Daytime Convection RGB, Figure \ref{fig:3} shows the cross section of cloud hydrometeors, particularly cloud liquid and ice mixing ratios, at the same time and across the area of interest where the severe convection persisted. In particular, concentration of cloud hydrometeors above the FZL (denoted as 0 $^{\circ}$C isotherm) and within the HGZ undergoes a crucial microphysical conversion or phase change into ice and mixed phase particles \cite{58}. The high concentration of these ice phase hydrometeors ($>$ 0.10 g kg$^{-1}$); nearly coinciding along the bulk of supercooled liquid contents, within the mid-level section of the storm's updraft favored growth of the droplet, collectively known as “embryos” - taking the form of graupel or frozen drops, by deposition and riming processes where the hailstone formation depends \cite{59}. Also, seen in the cross section, the presence of ice aggregates in the anvil of the convective storm ($>$ 200-hPa) validates the Daytime Convection RGB scan from HIMAWARI-8 where presence of ice particles was conformed and inline to the results of Gayet et al. \cite{60} and Stith et al. \cite{61}.

In addition, synoptic analyses revealed that a weak low pressure (more of an open wave) system was evident along the Philippine Sea, with minimal belt of wind flow rounding the 500-hPa low axis towards the Eastern Visayas region, mainly with northeasterly wind component seen in Figure \ref{fig:4}a. On the other hand, some easterly component were also seen and approached central Luzon, thus spreading of the wind vectors may indicate semblance of diffluent flow in the upper levels which aided on storm initiation. Above the PBL shown in Figure \ref{fig:4}b, the area was far removed from the aforementioned low pressure system as 850-hPa winds were accompanied by minimal magnitude along the area of just 5 kt. However, the case study was potentially located on a convergence boundary between the low-level winds from the southwest entering along western Luzon, including Metro Manila, and easterlies crossing the SMMR at higher levels. 

\begin{figure}[!ht]
    \centering
    \animategraphics[width=\textwidth, loop, autoplay]{1}%frame rate
    {sounding_gif/fig-}%path to figures
    {1}%start index
    {4}%end index
    \caption{00 UTC Tanay, Rizal Proximity and 06-08 UTC ERA5 Model Sounding Profiles. Legends are included in the upper-left corner of the figures. Annotations to the thermodynamic profiles include the SBLCL, PBL, MULFC, FRZ, and MUEL.}
    \label{fig:prof_gif}
\end{figure}

The interaction of these contrasting wind regimes at 850-hPa, particularly over Central Luzon, likely contributed to localized surface convergence. This may have facilitated low-level uplift, providing an initial trigger for convective development. Figure \ref{fig:4}c highlights the 0-3 km moisture flux convergence, where zones of moisture convergence align with the area of interest in Bulacan. Although the VIMFC was kept at minimal, the presence of these convergence zones (negative) suggests that moisture was actively being channeled upward, fueling the storm system. In fact, the storm initiation straddled and took place between a convergence area in the SMMR and area of sinking motion (positive) to the west. As indicated, these were at minimum, but Waldstreicher \cite{62} found that the favoured area of convection was located in the gradient between these areas. Moreover, the convergence boundary at lower levels, combined with the diffluent flow aloft, likely enhanced the vertical velocity fields within the column. This vertical motion is crucial in sustaining the deep convection and updrafts capable of lifting hydrometeors past the FZLs for hailstone growth \cite{63}.

\subsection{Mesoscale Environment}

\begin{figure}[!ht]
    \centering
    \animategraphics[width=\textwidth, loop, autoplay]{1}%frame rate
    {hodo_gif/fig-}%path to figures
    {1}%start index
    {4}%end index
    \caption{Associated with Figure \ref{fig:prof_gif} but for Ground-relative and Right-moving Hodographs. Annotations to the wind profiles include the individual Bunker's Storm Motion, Deviant Tornado Motion, and Corfidi's MCS components.}
    \label{fig:hodo_gif}
\end{figure}

\begin{figure}[!ht]
    \centering
    \includegraphics[width=\textwidth]{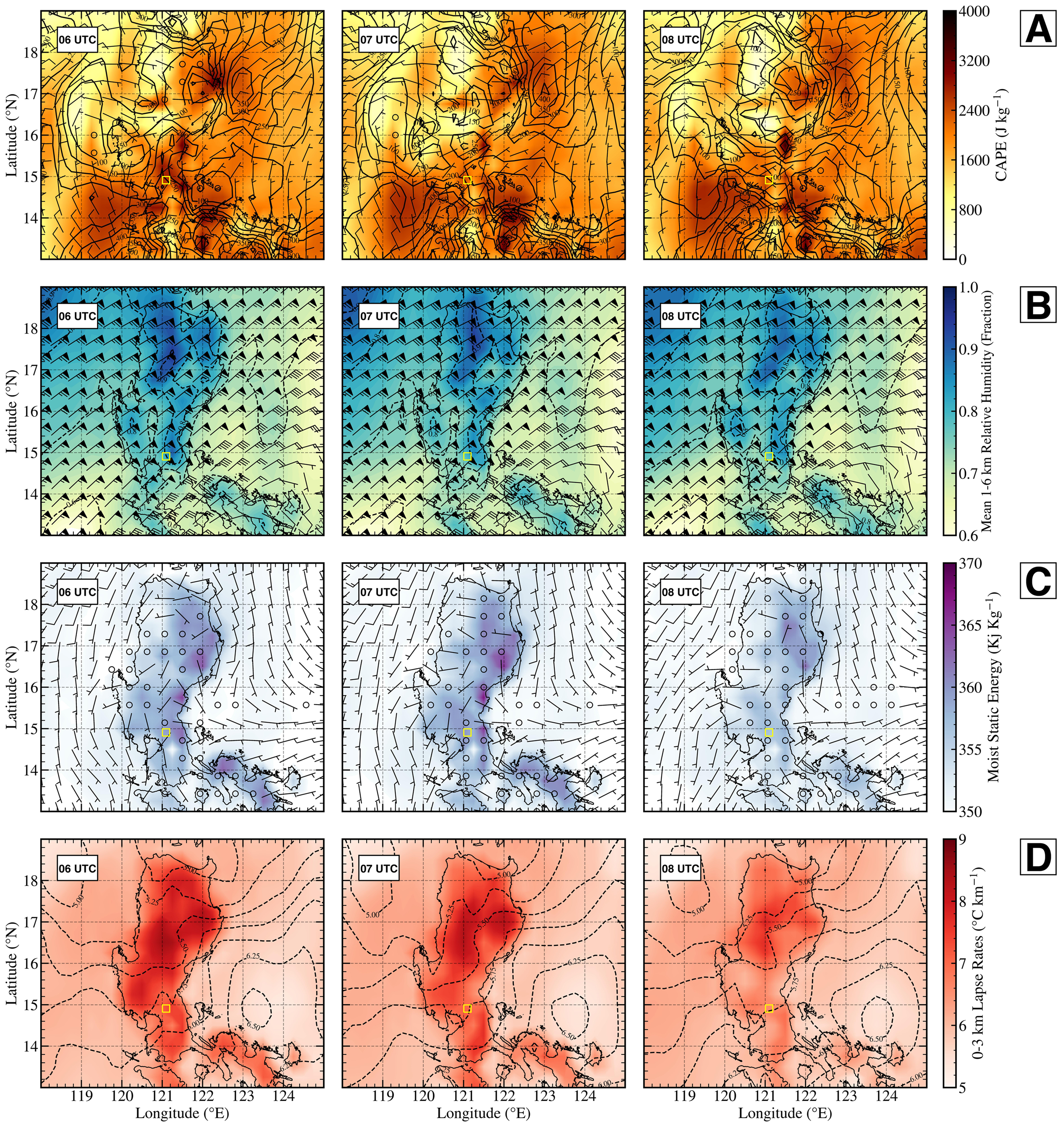}
    \caption{Convective and Kinematic environment of the Luzon landmass before (06 UTC), during (07 UTC), and after (08 UTC) the event. (a) CAPE (J kg$^{-1}$), 0-6 km Bulk Shear (kt), and WMAXSHEAR (black; m$^{2}$ s$^{-2}$). (b) Average 1-6 km Relative Humidity (fraction), including Average 1-3 km Relative Humidity (dashed; fraction), and Bulk Shear between the LFC-EL (kt). (c) Moist Static Energy (MSE; Kj kg$^{-1}$) and 10 m wind barbs (kt). (d) Contours of 0-3 km Lapse Rates ($^{\circ}$C km$^{-1}$) and 3-6 km Lapse Rates (dashed; $^{\circ}$C km$^{-1}$). The area of interest is demarcated in yellow box.}
    \label{fig:7}
\end{figure}

Pre-convective environment was shown and described in Figure \ref{fig:prof_gif} and \ref{fig:hodo_gif} through the Tanay, Rizal sounding at 00 UTC, a close proximity profile. The aforementioned figures will also include the ERA5 sounding profiles to be discussed later in this subsection.

The initial condition was characterized by minimal dewpoint-temperature spreads indicative of moist PBL and potentially low storm bases. At top, the profile diverges into a dry layer capping off the atmosphere (SBCIN of -14 J kg$^{-1}$) until sufficient convective heating and low-level moisture transport was present. This dry layer resembles to an elevated mixed layer (EML) from the desert southwest (i.e. great plains setup), typically accompanied by steep lapse rates as seen in the first 1 km, LFC located above the inhibition layer, and average RH between 1-3 km of 67\% just above the cloud base. In addition to the fairly well-mixed above PBL conditions, the environment was also conducive for effective downward momentum transfer and strong surface winds with initial DCAPE of 741 J kg$^{-1}$, potentially hinting outflow-dominated and wind-producing storms.  

\begin{table}[h!t!]
\small
\caption{Sounding-derived measurements associated for the hailstorm event.}
    \centering
    \begin{tabular}{lllll}
    \hline\hline
    Parameter & 00 UTC Obs & 06 UTC ERA5 & 07 UTC ERA5 & 08 UTC ERA5 \\
    \hline
    CAPE & 2081 J kg$^{-1}$ & 4663 J kg$^{-1}$ & 5029 J kg$^{-1}$ & 4044 J kg$^{-1}$ \\
    CAPE$_{\text{HGZ}}$ & 1000 J kg$^{-1}$ & 1776 J kg$^{-1}$ & 1901 J kg$^{-1}$ & 1630 J kg$^{-1}$ \\
    CAPE$_{\text{$<$HGZ}}$ & 619 J kg$^{-1}$ & 1117 J kg$^{-1}$ & 1167 J kg$^{-1}$ & 962 J kg$^{-1}$ \\
    ECAPE & 919 J kg$^{-1}$ & 2283 J kg$^{-1}$ & 2435 J kg$^{-1}$ & 2055 J kg$^{-1}$ \\
    DCAPE & 741 J kg$^{-1}$ & 663 J kg$^{-1}$ & 578 J kg$^{-1}$ & 608 J kg$^{-1}$ \\
    CIN & -14 J kg$^{-1}$ & 0 J kg$^{-1}$ & 0 J kg$^{-1}$ & 0 J kg$^{-1}$ \\
    LCL & 281 m & 762 m & 744 m & 636 m \\
    LFC & 1025 m & 762 m & 744 m & 636 m \\
    LR$_{03}$ & -5.24 $^{\circ}$C km$^{-1}$ & -6.91 $^{\circ}$C km$^{-1}$ & -6.96 $^{\circ}$C km$^{-1}$ & -6.45 $^{\circ}$C km$^{-1}$ \\
    LR$_{36}$ & -6.05 $^{\circ}$C km$^{-1}$ & -5.84 $^{\circ}$C km$^{-1}$ & -5.73 $^{\circ}$C km$^{-1}$ & -5.78 $^{\circ}$C km$^{-1}$ \\
    RH$_{13}$ & 0.67 / 67 $\%$ & 0.72 / 72 $\%$ & 0.70 / 70 $\%$ & 0.68 / 68 $\%$\\
    RH$_{16}$ & 0.59 / 59 $\%$ & 0.77 / 77 $\%$ & 0.75 / 75 $\%$ & 0.75 / 75 $\%$\\
    SHR$_{01}$ & 7 kt / 3.9 m s$^{-1}$ & 4 kt / 2.0 m s$^{-1}$ & 3 kt / 1.9 m s$^{-1}$ & 3 kt / 1.9 m s$^{-1}$ \\
    SHR$_{06}$ & 3 kt / 1.9 m s$^{-1}$ & 8 kt / 4.1 m s$^{-1}$ & 8 kt / 4.1 m s$^{-1}$ & 7 kt / 3.9 m s$^{-1}$ \\
    SHR$_{\text{LCL-EL}}$ & 69 kt / 35.5 m s$^{-1}$ & 60 kt / 30.8 m s$^{-1}$ & 57 kt / 29.3 m s$^{-1}$ & 62 kt / 31.8 m s$^{-1}$ \\
    SRW$_{01}$ & 14 kt / 7.2 m s$^{-1}$ & 13 kt / 6.7 m s$^{-1}$ & 13 kt / 6.7 m s$^{-1}$ & 13 kt / 6.7 m s$^{-1}$ \\
    WMAXSHEAR & 122 m$^{2}$ s$^{-2}$ & 395 m$^{2}$ s$^{-2}$ & 411 m$^{2}$ s$^{-2}$ & 350 m$^{2}$ s$^{-2}$\\
    WMAXSHEAR$_{06}$ & 67 m$^{2}$ s$^{-2}$ & 194 m$^{2}$ s$^{-2}$ & 198 m$^{2}$ s$^{-2}$ & 171 m$^{2}$ s$^{-2}$ \\
    FZL & 4190 m & 4665 m & 4725 m & 4715 m \\
    \hline
    \end{tabular}
    \label{table:2}
\end{table}

Furthermore, moist mid-level profile was also depicted with \textit{T}-\textit{Td} spreads converging just past 600-hPa with RH $>$ 80\%. However, the average tropospheric RH between the storm base and mid-level section of the storm remains dry with RH of just 59\% hinting potential entrainment \cite{64}. Close to 400-hPa, a 'warm nose' was also evident steepening the mid level lapse rates to within -6.05 $^{\circ}$C km$^{-1}$. This results to ample undiluted instability of 2081 J kg$^{-1}$, with CAPE$_{\text{HGZ}}$ of 1000 J kg$^{-1}$. As shown in the associated hodograph through Figure \ref{fig:hodo_gif}, combined with the weak storm-relative winds in the first 1 km, low-level shear value, and overall hodograph shape (straight line), the presence of unidirectional speed shear in the total column, particularly the LFC-EL shear magnitudes of 69 kt (35.5 m s$^{-1}$), compensated for the lack of 0-6 km bulk shear allowing convection to attain severity, be organized, and to initiate in the afternoon.

In addition, the mesoscale environment for this hailstorm event was viewed in the lens of ERA5 through Figure \ref{fig:7}. In this case, depicted in Figure \ref{fig:7}a, CAPE measurements of between 2000-2500 J kg$^{-1}$ was depicted suggesting ample instability in the atmosphere. However, DLS represented by the 0-6 km bulk shear were quite weak for organized storms ($<$ 5 m s$^{-1}$). Despite the ample instability from ERA5, this also resulted to weak combined WMAXSHEAR parameter of $<$ 300 m$^{2}$ s$^{-2}$ throughout the time unrepresentative of the storm's severity. In such cases where high CAPE and weak shear regimes are viewed, they are potentially linked to damaging winds within microbursts especially during warm-season severe weather periods \cite{65}. In Figure \ref{fig:7}b, sufficient Moist Static Energy within the virtual parcel, especially at 07 UTC approaching 360 Kj kg$^{-1}$ coupled with convective heating, allow thunderstorms to initiate along the subtle boundary formed by the opposition of low-level southwesterly winds and easterly turning to northeasterly winds aloft. Although the DLS vector was weak, thunderstorms were able to attain severe status due to the ambient cloud-layer bulk shear within the range of 55-60 kt from 06 to 08 UTC seen in Figure \ref{fig:7}c, with its shear axis being perpendicular to the initiating boundary owing its semi-discrete nature. 

Figure \ref{fig:7}d shows the temperature lapse rates within the Luzon from 06-08 UTC. The total instability one of the fuel for this storm cell, was the result of the steep temperature lapse rates of $>$ -6 $^{\circ}$C km$^{-1}$ in the lowest 3 km layer enabling warm-buoyant air parcels to precipitously rise and build instability within the area of interest. Given such low-level temperature gradient with lapse rates of $>$ -5.5 $^{\circ}$C km$^{-1}$ in the mid-troposphere, CAPE representations as discussed earlier may have been indicative that the thermodynamic instability can be larger than what is represented on these ERA5 spatial distributions. As a support this statement, for this case, ERA5's low-level and mid-level RHs shows the moistening of the environmental profile with RH $>$ 70$\%$ (Fig. \ref{fig:7}c). The combined presence of quality moist environment and mid-level instability results to more CAPE (so as sufficient buoyancy) to sustain the convective updraft \cite{66}, as compared to the pre-convective setup where RHs are $<$ 60$\%$. Thus, the standard CAPE in Figure \ref{fig:7}a should be treated with caution. 

To overcome the limitation, we combined the single and pressure levels from ERA5 by making the near-surface pressure, temperatures, dewpoints, heights, wind components as its initial starting point of the model sounding profile, rather than the standard procedure (start at the 1000-hPa or $\sim$110 km AGL). By re-gridding and re-computing the parcel profile, Figure \ref{fig:prof_gif} further depicts the derived model sounding at 06 UTC and it showed a well-mixed profile without any sign of inhibition layer, near-saturated PBL with RH $>$ 70$\%$, and accompanied by both steep low-level, especially before and during the hail event close to -7 $^{\circ}$C km$^{-1}$. This contributed for the warm-buoyant air parcels to substantially rise lifting hail-producing hydrometeors and generating excessive undiluted MUCAPE of $>$ 4000 J kg$^{-1}$ (at 07 UTC, it reached 5029 J kg$^{-1}$) more than adequate fuel for surface-based storms \cite{67}. 

In Figure \ref{fig:hodo_gif}, additionally, the wind profile associated with the sounding through the hodograph exhibited characteristics favorable for organized convection, including fairly unidirectional speed shear as a result from the zonal northeasterly flow. As seen in previous figures, the DLS is weak throughtout the event, but the LFC-EL shear $>$ 30 m s$^{-1}$ is adequate for storms to organize. In addition, SRW$_{01}$ were meager resulting to entrainment-driven dilution to the convective storm's updraft represented by a fractional entrainment of 48$\%$ was computed in the profile. This suggests that despite the moist profile, especially atop of the cloud base, the updraft of this storm is diluted and often only realize nearly half of the undiluted CAPE as updraft kinetic energy \cite{67a}. Still, the large CAPE ($>$ 4000 J kg$^{-1}$) compensated for the lack of DLS with WMAXSHEAR exceeded 400 m$^{2}$ s$^{-2}$ near the time of hailstorm event, well within the climatological bounds favorable for severe thunderstorms with hail threat \cite{23}. However, what further made the storm persistent, quite large spatially, and hail-producing can be attributed to internal and external storm dynamics. The external case was discussed later in the next subsection. Despite the entrainment effects and presence of excessive sub-HGZ CAPE of $>$ 1000 J kg$^{-1}$, Nixon et al. \cite{6} asserted that a balance between CAPE and SRW$_{01}$ exists favorable for hail producing environment. Inline to their results, our profile condition where both sufficient CAPE exists in the entire atmosphere and HGZ, including lower FZL, while at the presence of weak SHR$_{06}$ and SRW$_{01}$ is still favorable for hail growth along the updraft.

\subsection{Storm evolution from Satellite Perspective}

\begin{figure}[!ht]
    \centering
    \includegraphics[width=\textwidth]{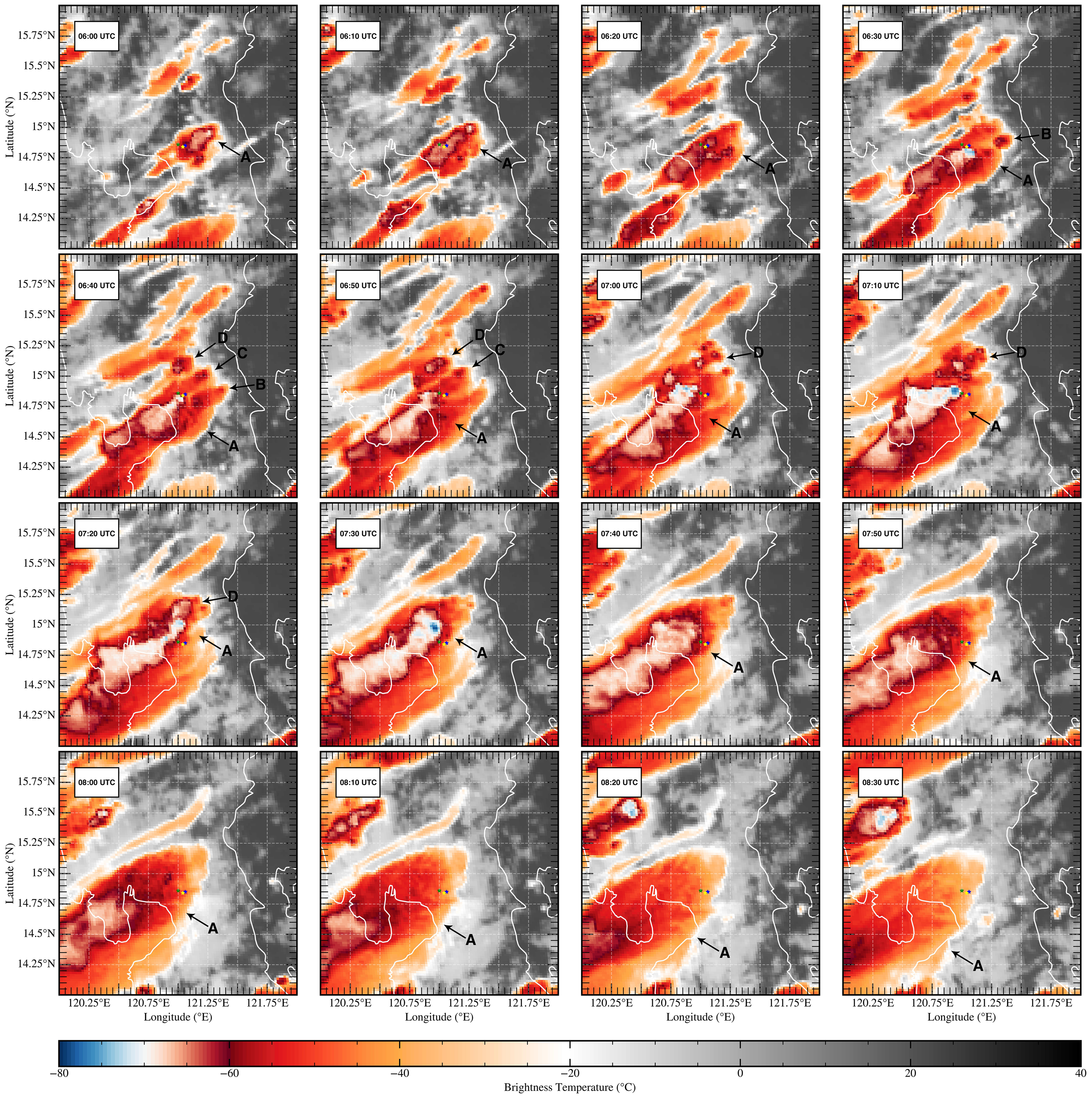}
    \caption{HIMAWARI-8 AHI 10.4 {\textmu}m BT depicting storm evolution of storm cells A (Main Storm), B, C, and D every 10 minutes, starting at 0600 to 0830 UTC. Colored stars are the impacted areas of Norzagaray (Blue), City of San Jose Del Monte (Yellow), and Sta. Maria (Green). }
    \label{fig:8}
\end{figure}

The evolution of the hail-producing severe storm from a satellite perspective is shown in Figure \ref{fig:8} from 0600 to 0800 UTC of HIMAWARI-8 on a 10-minute interval scans. However, prior to 0600 UTC, satellite images are not included in the visualization. Still, added in the descriptive context of the storm's life cycle.

\begin{figure}[!ht]
    \centering
    \includegraphics[width=\textwidth]{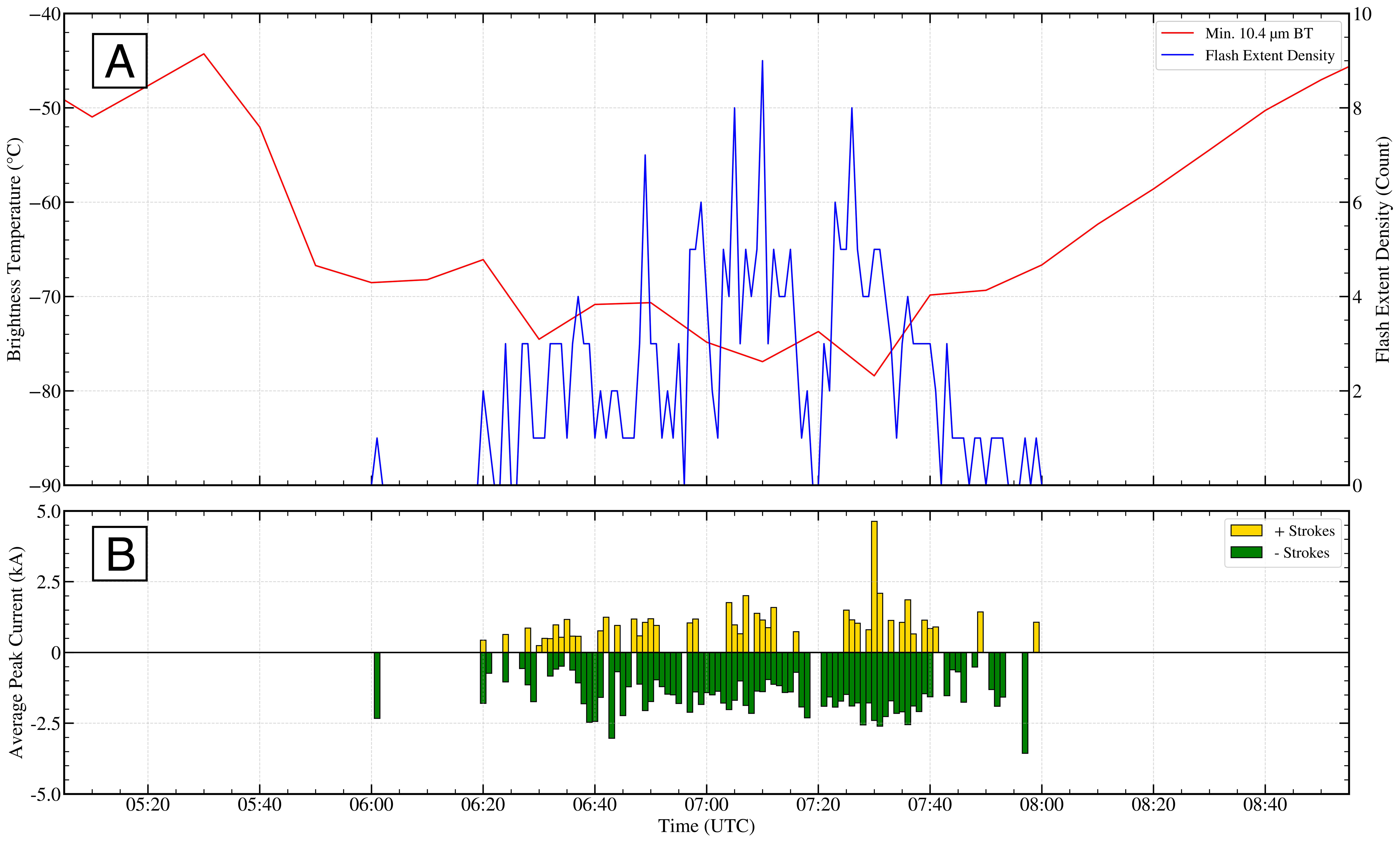}
    \caption{Time series of; (a) Minimum 10.3 {\textmu}m brightness temperature in the storm (red; $^{\circ}$C), maximum flash extent density per minute (blue; counts), and (b) Average Peak Current (kA) per minute of recorded Positive (yellow) and Negative Strokes (green).}
    \label{fig:9}
\end{figure}

The initial observed convective situation (04 UTC; not shown in Figure \ref{fig:8}) was described with cumulus field along the SMMR near the province of Bulacan as a byproduct of on-going surface heating and upslope flow from the near-surface southwesterly wind and from the easterlies aloft resulting to a small-scale convergence area. This convective situation is supported by the \href{https://www.facebook.com/groups/NCR.PAGASA/posts/4345762258804378/}{initial thunderstorm advisory} by DOST-PAGASA and its Regional Division located in the National Capital Region (PAGASA-NCR PRSD) issued at 0404 UTC where none was issued within the area of interest. Almost an hour after, activity and initiation ramped up with a single cell at its initial stage located at the higher terrain along a decaying cirrus plume to its south. The \href{https://www.facebook.com/groups/NCR.PAGASA/posts/4345865892127348/}{2nd thunderstorm advisory} by the PAGASA-NCR PRSD just 10 minutes after it was seen in the HIMAWARI-8 imagery asserts that storm activity has increased within Norzagaray area and its adjacencies with potential for moderate to heavy rainshowers at that time. 

By then, storm maturity is underway between 0540 to 0550 UTC scans of HIMAWARI-8 which will be our prominent storm (hereafter storm "A") seen in Figure \ref{fig:8}. By 0600 UTC, a discrete, strong TSTM developed at the slope of the terrain east of Norzagaray, Bulacan. 10 minutes after, PAGASA-NCR PRSD issued its \href{https://www.facebook.com/groups/NCR.PAGASA/posts/4345999962113941}{3rd thunderstorm advisory} citing most of the CALABARZON Region (i.e. Region IV-A) and parts of Central Luzon such as the provinces of Nueva Ecija, Tarlac, Zamboanga, and Bulacan; particularly, Dona Remedios Trinidad and Norzagaray are currently experiencing moderate to heavy rainshowers accompanied by frequent lightning and outflow winds due to the on-going deep convection that can persist over the next 2 hours. This storm propagated in an south-eastward direction due to the orientation of the cloud-level wind shear vector as discussed before. A second cell (hereafter cell “B”) initiated at 1830 UTC west of storm “A” but quickly dissipated minutes after. 

However, at least two auxiliary cells to the north of the main storm also emerged (hereafter cell "C" and "B") at around 0640 UTC. It is possible that both storm "A" and cell "D" interfered with the complete initiation of cell "C" by cutting off its moisture source, leading to its dissipation, while cell "D" remains as a sole auxiliary cell in close proximity with storm "A". In addition, Figure \ref{fig:8} also highlights a mature storm at 0710 UTC with cold cloud tops indicative of overshooting top structure above the anvil. While the secondary cell remains intact and eventually, at 0720 UTC, merges with storm "A" wherein another overshooting top feature is presented in its succeeding scan. The timing of the occurrence of this convective overshooting aligns with various hail reports between 0710 to 0720 UTC, thus the hailstorm lasted approximately 10 minutes within Norzagaray, Bulacan and nearby municipalities. Furthermore, the storm dynamic context as discussed also matches recent studies that most of the parent storms (both hail- and tornado-producing storms) are accompanied by auxiliary cells that eventually nudges and merges, constructively \cite{7}. This interaction may have played a role as well in countering the entrainment effects by providing ample ventilation and moistening of the upper-level profile. Eventually, storm "A" transitioned into an outflow dominant storm and dissipated past 0900 UTC.

The results shown in Figure \ref{fig:9} also indicate that the lightning flash counts from PLDN has some skill as a forecasting tool, given that the peak in electrical activity detected precedes storm intensification, both in terms of its minimum BT and emerging maturity of the storm cell. However, caution is advised when attempting to interpret the existence (or lack) of lightning jumps the PLDN. Little is known about the electrical characteristics of the storms that affect the province of Bulacan, and although the results shown here are promising, further research based on a larger sample of severe and non-severe storms is needed to properly evaluate the capabilities of using PLDN as a proxy for severe hail detection in the region. In general, the temporal evolution of the storm characteristics shows some resemblance to those found in Wapler \cite{68} for their analysis of severe hailstorms in Germany. This includes the increase in pulsing lightning activity prior to cell intensification and to the hail reports at the surface. The field measurement also depicts that about 78$\%$(22$\%$) of the lightning flashes in this storm are delivered through negative(positive) charges, inline to the results of Wang et al. \cite{68aa} for summertime thunderstorms. In addition, the escalation of lightning activity coincides with the substantial increase of WMAXSHEAR results compared to the initial sounding. It was found out that lightning increases with increasing values of WMAXSHEAR, especially in high shear environments \cite{23}. For this case, however, the inflation of instability: including microphysical processes i.e. mixed-phase ice nucleation-multiplication and rimming processes as previously discussed \cite[depicted in Fig. \ref{fig:3};][]{58,59,68a,68bb} and potential storm interaction \cite{7}, compensates for the lack of SHR$_{06}$ which can considerably increase the severe storm's lightning activity along the temporal domain of the event. 

\section{Conclusion}

This case study and data collected allows us to provide a characterization of the storm evolution and environment, which together with the multiple hail reports archived from Project SWAP are very useful to understand the different processes that act to modulate the storm initiation and severity. Synoptic analysis reveal that the severe thunderstorm may have formed between the areas of convergence and divergence as weak low-level flow came from the southwest and winds aloft with easterly component. Analysis of the pre-convective environment reveals that the development of an upslope flow, in combination with the eroding of the cap by means of the diurnal heating and subtle moistening low-level and mid-level troposphere, provide the conditions for convective initiation over the slope of the terrain as seen in later ERA5 sounding profiles. The thermodynamic profile transitioned from inhibited instability above the PBL to uncapped environmental profile with steep temperature gradients, especially in the first 3 km, resulting to excessive undiluted CAPE and lower FZLs. Although the DLS component were weak throughout the time, the presence of significant LCL-EL bulk shear, increased CAPE, and even storm nudging-merging process from an auxiliary cell at its matured life cycle, as the storms move away from the higher terrain favors the thunderstorm sporting a wider and stronger updraft that supports severe hail formation.

The combination of the available in-situ measurements and reports shows that hail is observed at the peak stage of the storm's life cycle (0710 UTC) where significant hail of diameter up to 5-8 cm is supported and lasted for about 10 minutes in the area of Norzagaray-City of San Jose Del Monte, Bulacan. The peak life cycle of the severe storm was diagnosed through both satellite and model analysis from ERA5 and reveals the presence of sufficient ice particles in the storm's updraft (within the HGZ) and anvil. We were not sure whether this severe storm is a supercell or not given that we do not have an access to the nearest radar, but close inspection of lightning flash counts showed that this transition is preceded by a significant increase in lightning activity, suggesting that this product has some potential for its use in lightning jump algorithms that could be used as a nowcasting tool. The storm also developed an overshooting top signature that was detectable through analyzing the BT signature previous to the time of the most hail. Identifying overshooting top and its cloud-top temperature also proves to be an effective tool in delineating the more severe phase of the storm, although it fails at pinpointing the occurrence of hail in the early stages of the convective storm.

Drawing general conclusions from a single case study is a difficult task. However, the opportunities in which it is possible to sample a storm in the region with the number of observational platforms that were available during the SWE is rare. Furthermore, as mentioned earlier, there are no available radars at that time (important for radar hail detection and convection mode characterization) and hence the analysis of the 'Friday the 13th' Bulacan Hailstorm is novel too in this sense. It is for these reasons that we think that it is important to present these kinds of observations that are unique and hopefully to be foreseen to be repeated, with better data coverage, in the near future within the Philippines.

Also, the forecasting and post-analysis techniques presented in this paper, while not a novelty, have not been tested in storms in this particular region. At present, despite the recent advances, the reliable prediction of hail by numerical weather prediction (NWP) models is still a challenging task, particularly because the microphysical processes involved are yet to be sufficiently understood, and are intimately linked with storm dynamics \cite{68b,68c}. It is also true that the coexistence of an increase in the observed lightning activity or an overshooting top signature in association with the largest hail in this particular storm does not mean that other severe hail storms may behave similarly. In addition, although recent research \cite[e.g.][]{69} shows that the environmental conditions in which hailstorms develop share some similarities in some regions of the world, those conditions may vary significantly even within those regions, making it difficult to generalize the results from a single case study. Parameters that are related to the convective updraft's strength have proven to be popular candidates in both Europe and the US owing to the nature of strong updrafts being necessary for hail growth \cite[e.g. CAPE, mid-level temperature lapse rate, bulk shear, storm inflow;][]{69a,69b} and even the depth of the HGZ parameterized as FZL \cite{69c}. Caution is advised as this approach remains limited due to the microphysical processes that impact the growth rate of hail and the role of storm mode \cite[][, and references therein]{69d}, both of which cannot be inferred solely from proximity soundings with a reasonable level of confidence. Still, we think that a case study with the number of observations presented and sounding parameters evaluated here is a good starting point from where much more extensive studies, some of them ongoing, may benefit.

The experience leads us to believe that the case study presented here is fairly representative of the severe storms that repeatedly affect the associated region (Region III - Central Luzon), including nearby regions such as National Capital Region (NCR), Region IV-A (CALABARZON). Those storms, like the one presented here, are often severe. The recent Project SWAP that pursued both active and inactive approach to collect hail reports from the general public to create a climatology of SWEs \cite{10} and a rigorous, upcoming hail study by Manalo et al. (\textit{in preparation}) shows that from 2007 to the half of 2024, hail season in the Philippines starts from May and clears off between September and October, inline to previous statements by Selga \cite{9}. The 'Friday the 13th Hailstorm' took place on midway of the hail season and is located to the identified hail hotspots. 

In addition, a recent growing idea on severe weather setups in the Philippines is the existence of an 'easterly severe weather setups' such as this case. Shown in the hodograph evolution figure, the winds aloft shifted to more (north)easterly component, while near-surface winds remain meager, thus the presence of unidirectional speed shear (or in storm lingo; a straight line hodograph or increasing speed shear with lack of 'turning'). Although we know already the initial characteristics of a southwesterly case e.g. raindrop size, amount of rainfall, temperature etc. \cite[e.g.][, and references therein]{70,71}, and even similar cases whose combined instability-kinematic properties resemble to the U.S. near the gulf of mexico \cite[GOM;][]{72}, such easterly setups can be attributed to monsoonal breaks, but its thorough characteristics relevant to storm severity are remain to be studied. A climatology of severe weather environments, without relying on rainfall, temperature, or any other near-surface variables, but on the instability profiles, buoyancy, entrainment, storm depth and inflow, maximum parcel height, and amount of shear across the entire column and specific layers is an acceptable way to diagnose such cases, not only for easterly setups, but also for classical southwesterly cases. 

For now, our next case study will be focused on tornadoes in the Philippines. First stop is the discrete tornadic supercell that erupted in Candating, Arayat in the Province of Pampanga back on 27th of May 2024 - a typical southwesterly setup for severe weather, 'typical' right?

\section*{Acknowledgments}

I am very grateful to the Philippine population for reporting the occurrence of hailstones in their communities. I also appreciate the valuable comments of the three anonymous reviewers and editor, which helped to improve this manuscript. This work received no funding, but was 'funded' by extensive and exhaustive effort, whose dedication and commitment to advancing our understanding of severe weather phenomena were indispensable. Finally, I am thankful to my family and my loved one for their unwavering support throughout this research.

\subsection*{Author Contributions} 
G. H. Capuli is the sole, corresponding author of this work. Thus, he conceptualized and leads the initiative, and designed-conducted the experiments/analysis. So as to the writing of the manuscript.

\subsection*{Conflicts of Interest}

The author declares that he has no competing interests.

\subsection*{Data Availability}
Data used in this paper were
derived from the ERA5 reanalysis (openly available through
the \href{https://cds.climate.copernicus.eu/#!/home}{Climate Data Source}), the individual full disk HIMAWARI-8 Bands were accessible through the  \href{https://thredds.nci.org.au/thredds/catalog/catalogs/ra22/satellite-products/arc/obs/himawari-ahi/himawari-ahi.html}{THREDDS Catalog}. The HIMAWARI-8 B13 colormap is customized and provided by M. P. A. Ibañez, a co-author for the upcoming Manalo et al. paper soon. The author's associated 1D vertical profile of standard ERA5 data and sounding data will be available soon through Project SWAP's next major release, but can be replicated using the location data through SounderPy at the lat-lon coordinates of (14.8$^{\circ}$ N, 121.0$^{\circ}$ E). The Digital Elevation Model (DEM) is from SRTM3 90m distributed and available in \href{https://portal.opentopography.org/raster?opentopoID=OTSRTM.042013.4326.1}{OpenTopography}. Finally, the lightning data from the PLDN is requested through PAGASA-CADS. Proper attribution is required for these datasets. This paper has made of use of the following Python packages:  \verb|Cartopy|, \verb|GeoPandas|, \verb|Matplotlib|, \verb|MetPy|, \verb|NumPy|, \verb|Pandas|, \verb|Rasterio|, \verb|rioxarray|, \verb|SciPy|, \verb|SounderPy|, and \verb|xarray|.

\printbibliography

\end{document}